\documentclass{article}
\usepackage[utf8]{inputenc}
\usepackage{url,hyperref,lineno,microtype}
\usepackage[onehalfspacing]{setspace}
\usepackage{natbib}     
\usepackage{graphicx}
\usepackage{tabularx}
\usepackage{amsmath}
\usepackage{amssymb}

\graphicspath{{imgsub/}}

\newcommand{\eq}[1]{\begin{equation}#1\end{equation}}

\newcommand{\eqa}[1]{\begin{eqnarray}#1\end{eqnarray}}
\newcommand{\avec}[1]{\left( \begin{array}{c}#1\end{array} \right)}
\newcommand{\avecc}[1]{\left( \begin{array}{c c}#1\end{array} \right)}
\newcommand{\mat}[1]{\mathbf{#1}}
\renewcommand{\vec}[1]{\boldsymbol{#1}}

\begin{document}

\begin{flushleft}
{\Large
\textbf\newline{Generalized Modeling: A survey and guide}
}
\newline
\\
Jana C. Massing\textsuperscript{1,2,3*},
Thilo Gross\textsuperscript{1,2,3},
\\
\bigskip
{1} Helmholtz Institute for Functional Marine Biodiversity at the University of Oldenburg (HIFMB), Ammerl\"ander Heerstr. 231, Oldenburg, Germany.
\\
{2}  Alfred-Wegener-Institute, Helmholtz Centre for Marine and Polar Research, Am Handelhaven 12, Bremerhaven, Germany.
\\
{3} Carl-von-Ossietzky University, Institute for Chemistry and Biology of the Marine Environment (ICBM), Carl-von-Ossietzky Str.~9-11, Oldenburg, Germany.
\\
\bigskip
* jana.massing@hifmb.de

\end{flushleft}

\section*{Abstract}
Many current challenges involve understanding the complex dynamical interplay between the constituents of systems. Typically, the number of such constituents is high, but only limited data sources on them are available. Conventional dynamical models of complex systems are rarely mathematically tractable and their numerical exploration suffers both from computational and data limitations. Here we review generalized modeling, an alternative approach to formulating dynamical models. We argue that this approach deals elegantly with the uncertainties that exist in real world data and enables analytical insight or highly efficient numerical investigation. We provide a survey of recent successes of generalized modeling and a guide to the application of this modeling approach in future studies such as complex integrative ecological models.

\section{Introduction}
Much of the power of mathematics comes from its ability to describe unknown objects. Consider the number $\pi$: almost everybody knows some of its digits, but nobody knows its exact value. Our ignorance of the value, however, does not impinge on its utility in calculations, nor does it prevent us from exploring its properties.

The ability of mathematics to work with unknown objects is not limited to numbers. In modeling we can exploit this ability, by writing models in terms of unspecified functions.
This is particularly useful in social or biological models where actual laws realized in nature are only approximately known.

In mathematics, working with unknown functions is the rule rather than the exception. For example, we do it every time we write the definition of a derivative. 
\begin{equation}
\dot{x}=\frac{{\rm d}}{{\rm d} t} x(t) = \lim_{\delta\to 0} \frac{x(t+\delta)-x(t)}{\delta}
\end{equation}
In models unspecified functions have been used to a lesser extent, but still the examples of such models are plentiful. 

In the present paper we examine the method of Generalized Modeling (GM), which is a formalism by which specific types of 
insights can be extracted from models containing unspecified functions. Although the approach has been extended to many different contexts, GM has mostly been applied to explore systems of ordinary differential equations (ODEs). For this class of systems, GM can provide insights into the dynamics, identify conditions for the stability of steady states, 
explore the response of steady states to perturbations, identify parts of the system that are particularly vulnerable to perturbations, etc.
Perhaps most importantly, GM generates these insights highly efficiently. It thus enables the
exploration of large complex many-variable systems with comparatively little effort.

In the following we explain the basic idea of GM in Sec.~2. After this philosophical introduction, we provide a comprehensive review 
of previous works that used GM in Sec.~3. This is done partly to point the reader to specific variants and applications that may suit their particular needs, but also partly to illustrate the power of GM, which played a crucial rule in several discoveries leading to high-profile publications. We then provide an introductory example to for the GM procedure in Sec.~4. In our experience, seeing this procedure leads to a set of specific questions, which we address in a frequently-asked-question section, Sec.~5. We then discuss the GM approach more extensively in Sec.~6, while paying particular attention to some decisions that need to be made during the modeling process. Different ways in which GM can be analyzed are then discussed in Sec.~7, before a concluding discussion in Sec.~8. 

\section{Basic Idea of Generalized Modeling}
GM is best understood by contrasting it against conventional stability analysis of steady states.
The conventional approach can be regarded as a 3-step process.
\begin{enumerate}
\item Parameterization: Restrict the model to equations that are specified except for a number of unknown parameters. 
\item Steady States: Find the steady states of the ODE.
\item Linearization: Compute the Jacobian matrix, which provides a linearization of the dynamics around the steady state. 
\end{enumerate}
Once the Jacobian has been obtained it can be used to explore the stability of steady states, find their bifurcations,
gain insights into non-stationary dynamics, etc. 

It is interesting to note that the three steps of this basic program above involve very different difficulties: The first is 
not technically difficult, but it involves ``artistic freedom'': it is easy to come up with some model but it may require much experience to find the best model for a given phenomenon, and sometimes it is not even clear what constitutes the best model. By contrast, the second step has a clear-cut answer, but we need to find the roots of an equation system, a task that is prohibitively difficult for all but the simplest systems. We are thus often forced to turn to numerics, but even then, no algorithms with guaranteed convergence are known. Finally, the third step only involves differentiation of functions, which is generally easy. In the worst case we can compute the derivatives by finite difference methods.

In terms of the actual technical difficulty step 2, the computation of steady states clearly stands out. In return for braving these 
difficulties we obtain the number of steady states and their locations. This can be valuable information for some systems, but in many other cases we know the steady states from observation of the system and thus the prediction from the model is often merely used to eliminate some unknown parameters by setting them such that the predicted states from the model match up with their observed counterparts.

Given that the computation of steady states can introduce significant difficulties in the modeling process, but reveals only limited information, it is interesting to ask if we can circumvent this step. For example random matrix models achieve this by directly formulating a model for the Jacobian matrix, rather than deriving the Jacobian matrix from ODEs \citep{wigner1955characteristic,May1972Nature,allesina2012stability}.
The power of random matrix models is illustrated impressively by Robert May's seminal work \citep{May1972Nature},
where he formulated a random matrix model for complex food webs. Exploiting the power of the random matrix May was able to 
prove mathematically that large random food webs are unlikely to be stable. The model thus proved decisively that the
large food webs observed in nature must have some special features that lends them there stability. At the same time the 
abstract random matrix formulation gave researchers very little intuition as to what these features might be.
Random matrix models are powerful because they give us direct access to Jacobian matrices in a sufficiently simple form 
to allow rigorous mathematical and highly efficient numerical exploration. However, as they lack the underlying layer of 
differential equations, these models tend to be more abstract and hence are often not easily interpretable. 

GMs are situated at the halfway point between conventional and random matrix models. They have almost the full power and efficiency
of random matrix models while being almost as interpretable as conventional ODE-based models. To understand how this is possible 
let us consider the three steps of the modeling process again. In GM we do not restrict the processes in the model to 
specific functional forms, and thus we cannot meaningfully compute the steady states of the model. This means in the GM 
the steady states are unknown quantities. However even though the steady states are unknown we can still formally linearize the dynamics around them, which yields Jacobian matrices. 
At first glance the elements of these Jacobian sound intimidating: They are derivatives of unknown functions in unknown steady states. However, these elements can be expressed in terms of a small set of parameters that have a clear and intuitive interpretation in the context of the model. 

In GM the three steps of the modeling procedure are thus reordered and slightly modified
\begin{enumerate}
\item Steady states: Consider a class of models that is general enough that steady states must exist in this class. Define symbols to denote the variables in these unknown steady states.
\item Linearization: Formally compute the derivative of the processes with respect to variables to compute the Jacobian.
\item Parameterization: Identify the quantities that appear in the Jacobian as parameters of the model. 
\end{enumerate}
The result is a prescription for generating the Jacobian of a steady state as a function of a number of possibly unknown but interpretable parameters. 
In other words, we directly get the Jacobian matrix in a steady state, which is reminiscent of a random matrix model. However, simultaneously our interpretation of this matrix profits from the underlying layer of differential equations almost as if it were a conventional model (see Tab.~1).

The heart of the GM procedure and the feature that sets GM apart from other models containing unknown functions, is the parameterization step, where we use a specific mathematical identity to give meaning to the parameters. This is explained in more detail in Section 4.

In the next section we are reviewing some of the past successes of GM. Readers who are eager to see an example of the procedure first may want to skip ahead to Section 4. 

\begin{table}
\caption{Comparison of modeling approaches}
\footnotesize
\begin{tabularx}{\textwidth} { 
  | >{\raggedright\arraybackslash}X 
  | >{\raggedright\arraybackslash}X 
  | >{\raggedright\arraybackslash}X 
  | >{\raggedright\arraybackslash}X |}\hline
Approach & Information needed & Most difficult step & Insights gained \\\hline
Conventional model (specified ODEs) & State variables and processes, specific functional forms for processes & Steady state computation or numerical simulation & Simulated time-series, number and stability of steady states, ...  \\
Random matrix model & Statistics of matrix elements & Analytical solution or eigenvalue computation & Stability conditions, nature of bifurcations \\
Generalized model & State variables and processes & Analytical solution or eigenvalue computation & Stability conditions, nature and location of bifurcations, response to perturbations, identification of interesting parameters and parameter regions
\\\hline
\end{tabularx}
\end{table}


\section{Generalized Models in the Literature}
Since its inception 12 years ago, GM has been applied to a wide range of subjects. In this section we review the areas where GM has made an impact and the ways in which the methodology has been adapted to suit to the various fields.

\subsection{Food Web Models}
The first GM was a simple predator-prey model proposed by Wolfgang Ebenh\"oh in 2003. An analysis of this model was eventually published in \cite{Gross2004PD}. The predator-prey model was subsequently expanded into a general food-chain model. Analysis of this model revealed minor details in the shape of the functions used in conventional food-chain models can have a strong impact on stability properties \citep{Gross2004JTB}, the same insight was discovered almost simultaneously in a different way in \cite{fussmann2005community}. A similar GM setting, exploring the so-called paradox of enrichment, was recently studied in \cite{awender2021stability}.

While GM was initially viewed as a trick that worked in one particular model, the subsequent extensions implemented in \cite{Gross2005OIKOS} made clear that the approach is generally applicable. This lead to an early paper that presented the GM as a general methodology \citep{Gross2006PRE}. This paper concluded by deriving a general food-web model, but did not analyze it in any detail.

Around 2006 GMs were still studied mostly by analytical computation of the bifurcations \citep{GrossThesis}. Although the bifurcation had been computed in food chains of up to ten levels, the structural complexity of food-web topology, still presented a serious obstacle. Instead GM approach was extended to predator-prey systems in space, modeled by partial differential equations \citep{Baurmann2007JTB} and was applied to study the effect of predator interference \citep{vanVoornMBE2008} and the dynamics of ecoepidemic models \citep{Stiefs2009MBE} and to explore the impact of nutrient content on predator-prey systems \citep{Stiefs2010AMNAT}. The latter paper resolved a controversy that arose because different previous models for nutrient content predicted very different bifurcation diagrams. Analysis of the GM showed that all of these diagrams were projections of the same bigger picture and identified the specific assumptions that explained the differences in the respective projections. 

\begin{figure}[ht]
    \centering
    \includegraphics[width=0.6\textwidth]{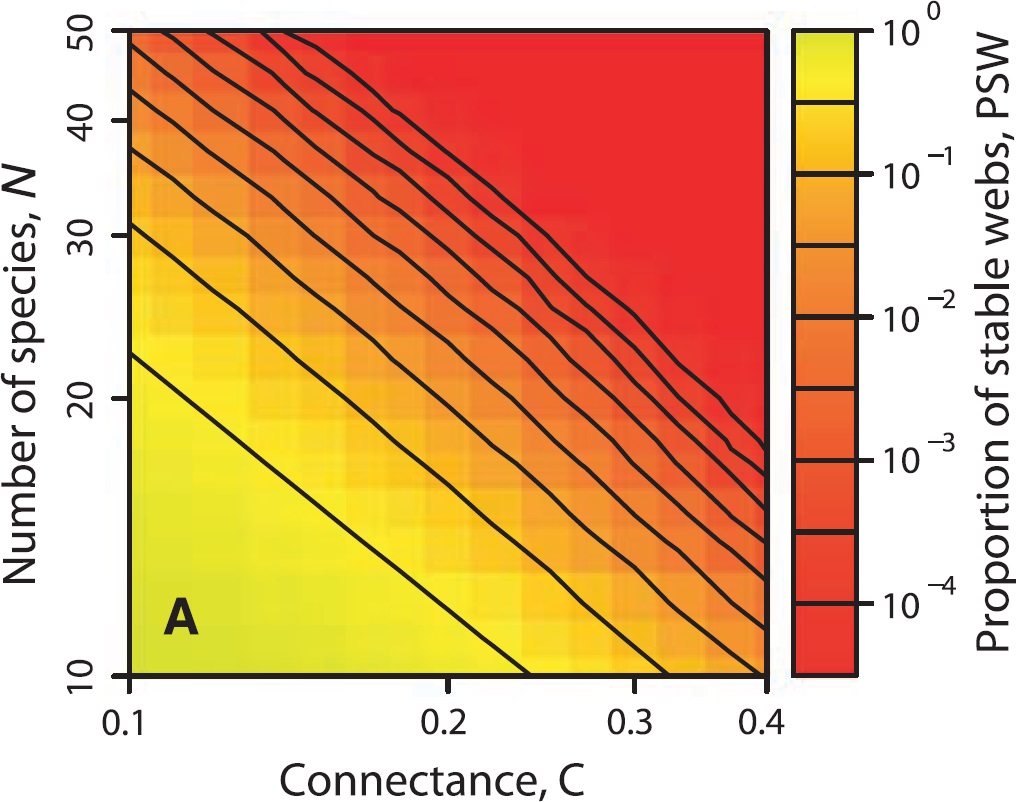} 
    \caption{Stability of complex food webs. Color coded is the probability that a steady state in a niche-model \citep{williams2000simple} food web is stable, PSW. The result confirms May's random matrix result that large complex food webs are typically unstable. To make this figure a GM was used to analyze 35 billion different networks. The paper in which it appeared used $10^{11}$ niche-model food webs. Figure reprinted from \cite{Gross2009Science}.}
    \label{figLarsplot}
\end{figure}

By 2009 work in metabolic models \citep{Steuer2006PNAS} had established numerical procedures for the investigation of GMs. This step provided an efficient method for the exploration of the food-web model formulated in \citet{Gross2006PRE}. The first application of this model focused on food-web stability. Since May's work, described above, identifying the properties that lend large food webs their stability had been a persistent challenge in ecology. Previous work had made progress by simulating systems of ODEs \citep{Williams2004EPJ,McCann2005EcolLett,Brose2006EcolLett,neutel2002stability}, however numerical limitations meant that only on the order of some thousand randomly generated food webs could be considered. By contrast, the higher numerical efficiency of the generalized food-web model allowed to study ca.~100 billion ($10^{11}$) randomly generated food webs within a month (Fig.~\ref{figLarsplot}). Building on this data several previous insights on food-web stability were confirmed, although evidence for the very popular weak link hypothesis \citep{McCann1998Nature} was only seen in smaller webs, instead a new topological pattern that contributed strongly to stability was identified \citep{Gross2009Science}. In response to these results there was a sharp increase in the interest in GM and the methodology was adopted by several labs.  

Barbara Drossel and coworkers carefully examined the generalized food-web model and its relationship to conventional food-web models. They were thus able to narrow down the ranges for generalized parameters and as a result found that this increased the proportion of stable states that were found in the food webs \citep{Plitzko2012JTB}.

Because generalized ecological models yield tractable Jacobian matrices even for relatively complex systems, they were used as a platform for a number of methodological developments. For example \citet{Stiefs2008IJBC} employed the approach to develop a method to visualize bifurcations, \citet{Lade2012PLOS} proposed a new type of warning signal for critical transitions based on GM and \citet{Hoefener2011EPL,Hoefener2012PTRS} used GM to study a delay-coupled network of populations. The latter work led to an algorithms for designing \emph{dynamic motif} small subgraphs of a network that exhibit specific dynamical instabilities regardless of the networks that they are embedded in \citep{Do2012NJP}.  

The GM approach was further refined in \citet{Yeakel2011TE}, which carefully examined the modeling procedure, and 
\citet{Kuehn2013AppMath} which first supported GM by rigorous mathematical work and then went on to extend the approach to the analysis on non-local dynamics \citep{Kuehn2013DCDS}.

Another extension is found in the work of Helge Aufderheide, who considered eigenvector localization in GM. He was able to explain why certain food webs have a different structure but the same generalized bifurcation diagram \citep{Aufderheide2012NJP}, a phenomenon first noticed in \citet{GrossThesis}. In a subsequent work eigenvector methods were used to propose an approach for identifying the species in a food web that are most susceptible to perturbations and those that have the strongest impact on the dynamics of the system \citep{Aufderheide2013PNAS,doizy2018impact}. 

Aufderheide's approach was subsequently used by Yeakel and coworkers to analyze a 6000 year time-series of Egyptian food webs \citep{Yeakel2014PNAS}. Yeakel had reconstructed an ensemble model of mammalian food webs from depictions in Egyptian art history. This dataset was then fed into the GM, which showed that extinctions from climate change events and the human population growth at the beginning of the 20th century reduced the stability of the web leaving it more and more vulnerable to further perturbations. It was also confirmed that vulnerable species, identified by GM in the initial network, were among the first to go extinct. 

The utility of the GM approach for combining complex social and ecological dynamics in a common social-ecological model was highlighted in \cite{Lade2015ArXiv}. Using GM, the authors explored the impact of human behavior on ecological systems \citep{Lade2013TheorEcol}, showing that social dynamics have a strong impact on tipping points. A subsequent review discussed the use of GM in socio-ecological systems more generally \citep{lade2017generalized}. Other environmental applications of GM include the analysis of a climate model by \citet{knopf2006multiparameter} and stock recruitment by \citet{yeakel2014generalized}. 

More recently GM was used by different labs to study meta-food webs, a class of models where food webs in different spatial patches are coupled by dispersal \citep{leibold2004metacommunity}. After initial works considered a single population model on a spatial network \citep{Tromeur2016JTB} and food webs on two patches \citep{Gramlich2016JTB}, the dynamics of complex food webs in large spatial networks were studied in \citet{brechtel2018master}. In this work methods from algebra were used to show that it is possible to derive master stability functions \citep{segel1976application,pecora1998master}, which then govern the food-web stability in any spatial network. 

\citet{anderson2021body} used GM to study the effects of positive body size scaling of dispersal on the stability of heterogeneous metacommunities. Their results cast doubts on the widely held opinion that the ability of large bodied predators to migrate farther than small bodied species is crucial for stability. 

\subsection{Models of Metabolism}
A second area where GM has been frequently applied is studies of metabolism. This line of work was started by Ralf Steuer in his PhD thesis. The metabolic version of GM is also known as \emph{structural kinetic modeling} after the title of his first publication on the subject \citep{Steuer2006PNAS}. In this paper Steuer and coworkers demonstrate that GM can be applied to metabolic systems such as glycolysis in yeast and the photosynthetic Calvin cycle. For the case of glycolysis it was shown that the GM approach could be used to exactly predict one of the parameters in the system, based on stability considerations. These and other findings were later confirmed in \citet{Gehrmann2011PRE}, who used GM to analyze a more complex model of glycolysis. 
\citet{carbonaro2017using} applied structural kinetic modeling to determine metabolic components that are major contributors to network stability in complex metabolic networks associated with glycolysis and pentose phosphate pathway and to predict the impact of perturbations on these components. 

In a subsequent paper \citet{Steuer2007BI} analyzed the TCA cycle in mitochondria (also reviewed in \citep{Steuer2007Phytochem}). To deal with this more complex network they proposed a numerical sampling procedure, explained in detail in the next section. This procedure allowed Steuer to explore the model efficiently. A specific biological question driving this research was why the mitochondria under consideration hardly utilized pyruvate as an energy source. This was resolved when the GM identified pyruvate import into the mitochondrion as one of the main drivers of instability.  

The sampling procedure of Steuer greatly increased the scope of GM and turned it into a highly efficient tool for the analysis of large networks. A matlab toolbox for metabolic GMs facilitating this analysis was published by \citet{Girbig2012BI}. The methodology was further refined by a careful exploration of various measures to reveal important regulators \citep{Grimbs2007MSB}. They combine GM and machine learning (ML) techniques to identify bifurcations in large systems \citep{Girbig2012Plos} and incorporate thermodynamic constraints \citep{Childs2015BI}. The review by \cite{Srinivasan2015BioTech} highlights the potential of this GM+ML approach to scale to whole cell models.

GMs were also used to study the inhibitory feedbacks in the sucrose cycle \citep{Henkel2011EURASIP} or in designing drugs. The latter 
is done by comparing two different metabolic states (e.g. the healthy and non-healthy system)
and their responses to perturbation with each other \citep{Murabito2011Interface, Murabito2013CSSB}. 
In addition, structural kinetic models are also used to set up more complex models of huge metabolic networks, i.e. hybrid models. 
Hybrid models describe central processes in high detail, while others
are roughly approximated. The central processes can be identified using structural kinetic models \citep{Bulik2009FEBS}.

\citet{Reznik2010JTB} applied GM in the analysis of further metabolic cycles and showed short cycles to be highly stable. This is particularly true for non-autocatalytic cycles \citep{Reznik2013JRSI}. 
Extending the methodology to metabolic genetic circuits showed that timescale separation between subsystems has a stabilizing effect \citep{Reznik2013Chaos}.
Subsequent works explored the dynamics of common regulatory motifs \citep{Zumsande2010JTB,Gehrmann2010PRE,Ackermann2012EPJ}, systems of interacting compartments \citep{Fuertauer2016TheoryBiosci}, and applied GM in a metabolic engineering application \citep{ye2015dynamics}. 

\subsection{Other applications and similar approaches}
In the medicine, GM were used to identify an early warning signal for critical transitions in systemic inflammation \citep{scheff2013predicting}. Moreover, \citet{zumsande2011general} proposed a GM of bone remodeling, leading to the identification of dynamical instabilities, which explain certain physiological and pathological dynamics of bones. 

GMs were also applied to study questions in social dynamics and management. \citet{Gross2006PRE} used a model of the Chinese Dynastic cycle as one of its examples, and the dynamics of manufacturing supply networks is analyzed in \citet{ritterskamp2018revealing, demirel2019identifying}.

A similar approach to GM is the analysis applied in \cite{Kisdi2013140}, that uses unspecified evolutionary trade-off curves to identify trade-off functions that lead to stable limit cycles in eco-evolutionary models. Another related method is the general structural sensitivity analysis proposed by \cite{Adamson2014JMB}, that considers the infinite-dimensional neighborhood of model functions to determine the sensitivity with regards to the local stability of steady states. Thereby, they provide a method to conduct bifurcation analysis under uncertainty in model functions and to determine probabilities of certain bifurcations \citep{Adamson2014BMB}. Their framework shares the use of unspecified functions, the steady-state treatment and the in-cooperation of the values of unknown functions as parameters in the Jacobian with the GM approach.


\section{An introductory example}
Consider a system where a variable, $X$, changes dynamically in response to gains and losses. We can write the differential equation 
\eq{
\label{eqGMex}
\dot{X}=G(X)-L(X),
}
where the dot denotes a time derivative, and $G$ and $L$ are unknown functions representing the gain and the loss terms. We have so far not constrained these functions in any way, so the only assumption is that gain and loss are in principle describable by mathematical functions. 

In conventional modeling we would now proceed by parameterizing the gain and loss functions, i.e.~restricting them to specific functional forms. Thereafter we could compute steady states and then launch into deeper analysis, computing stability, bifurcations etc. GMs build on the insight that most of these deeper analyses do not actually require us to restrict the processes to specific functional forms. 

For example the stability of steady states is captured by the so-called Jacobian matrix $\bf J$, defined as 
\eq{
    J_{ij} = \left.\frac{\partial}{\partial X_j} \dot{X}_i \right|_*
}
where we used $|_*$ to indicate that the expression is evaluated at the steady state under consideration. For our one-dimensional example system the Jacobian is a $1\times 1$ matrix and its only element is 
\eq{
J_{11} = G'(X^*)-L'(X^*)
}
where the dash denotes a partial derivative and $X^*$ is the steady state under consideration. Without further assumptions $G'(X^*)$ is the derivative of an unknown function evaluated at an unknown point. However, we know that terms such as $G'(X^*)$ represent numbers. We can therefore think of $G'(X^*)$ as an unknown parameter of the system. However, defined in this way, the parameter does not have an intuitive interpretation in the context of the application.

GM (in the narrow sense) is a particular way of parameterizing models such that we avoid restricting the processes to specific functional forms while capturing the uncertainty about the system in easily interpretable parameters. 

To parameterize the Jacobian in an interpretable way we need to make one more assumption: All variables and process rates have positive values. In many applications this is is very intuitive as variables describe quantities that are naturally non-negative, and process rates have positive values by design (e.g.~a gain would not be a gain if it were negative). The special case of one variable or process becoming exactly zero is discussed below. 

Let's return to the specific example of Eq.~\ref{eqGMex}. The equation describes a class of models in which positive steady states are bound to exist. This is not an additional assumption but merely the effect of working with a broad class of models rather than one particular parameterization. We denote these steady states as $X^*$ and denote the rates of processes in the steady state as $L^*=L(X^*)$ and $G^*=G(X^*)$, respectively. Although we use $X^*$ as a placeholder for every positive steady state in the system, we can formally normalize the equations with respect to $X^*$,
\eq{
x=\frac{X}{X^*},
}
such that $X=xX^*$. Likewise, we can define normalized
\eq{
g(x) = \frac{G(xX^*)}{G^*} \quad\quad l(x)=\frac{L(xX^*)}{L^*}
}
Here we followed the GM convention of using lower-case variables for normalized quantities and upper-case variables for unnormalized quantities. We can write a differential for the normalized variable as
\eq{
\dot{x}=\frac{\rm d}{{\rm d}t}\frac{X}{X^*}=\frac{\dot{X}}{X^*} = \frac{G(X)-L(X)}{X^*} = \frac{G^*}{X^*}g(x)-\frac{L^*}{X^*}l(x). 
}
By normalizing we have moved from a system in which we did not know the steady state to a system where we know it: In the normalized system the steady state is $x^*=1$ and in the steady state all processes run at rate 1. The price that we have to pay for this convenience is the appearance of the two factors $G^*/X^*$ and $L^*/X^*$. Because these factors are unknown scalars we can interpret them as unknown parameters of the system. 

Note that $G^*/X^*$ is the per-capita gain per X in the steady state. If X is a biological population we would call it the birth rate, and $L^*/X^*$ would be the per-capita death rate. Even in other systems such per-unit turnovers are generally well interpretable. 

Another interesting observation about the two parameters is that they must be equal as $G^*=L^*$ must hold in all steady states. This allows us to define
\eq{
\alpha = \frac{G^*}{X^*} = \frac{L^*}{X^*}.
}
Such parameters are known as \emph{scale parameters} in GM.
Our system now becomes 
\eq{
\dot{x}= \alpha (g(x)-l(x)).
}
The identity of the two fractions is a double-edged sword. We essentially used the stationarity property of the steady state to reduce the number of parameters, which generally leads to welcome simplifications. However, if we forget to exploit this simplification we may end up investigating steady states which cannot exist in the real world (say those with $G^*\neq L^*$). This is the one risk in GM that we need to steer clear of. Below we present a simple procedure for larger models that takes care of this point almost automatically. 

Having successfully normalized the model we are now ready to launch into the stability analysis of the steady state $X^*$. For our small example the Jacobian is 
\eq{
\label{eqexjac}
{\bf J} = \left[ \alpha (  g'(1) - l'(1) )    \right]
}
where the one appears because $x^*=1$. Since we still haven't constrained the functions $g$ and $l$ we don't know their derivatives, hence they are also unknown parameters of the system. We define
\eq{
g_{\rm x}=g'(1) \quad\quad\quad l_{\rm x}=l'(1)
}
To understand the interpretation of these parameters, consider what would happen if, say, the loss were a linear function, $L(X)=aX$, with $a>0$. In this case the normalization would result in $l(x)=x$, regardless of $a$, and hence $l_{\rm x}=1$. So every linear relationship results in a parameter value of 1. Furthermore, any power law, $L(X)=aX^p$ results in $l_{\rm x}=p$. So a quadratic relationship would be signified by a parameter value of $2$, a square root by $1/2$, and a reciprocal relationship, e.g.~$L(C)=a/X^p$ by a parameter value of -p. 

Parameters such as $g_{\rm x}$ and $l_{\rm x}$ are called elasticities. In the context of GM they are also called \emph{exponent parameters}. One can show that they are the logarithmic derivatives of the original functions. For example 
\eq{
l_{\rm x} = \left.\frac{{\rm d}\ln L }{{\rm d}\ln X }\right|_*.
}
We could have saved ourselves some work by using the mathematical identity
\eq{
\left.\frac{\partial L}{\partial X}\right|_* = \left.\frac{L^*}{X^*} \frac{\partial \ln L}{\partial \ln X } \right|_* = \alpha l_{\rm x},
}
however explicit normalization is generally felt to be the saner and safer way to a GM.

Elasticity parameters were originally introduced in economic theory
\citep{reilly1940use}, where they remain in wide use. In biology, elasticities are central parameters studied by metabolic control theory \citep{fell1985metabolic}. Besides their convenient interpretation, elasticities provide a measure of nonlinearity that can be very robustly estimated based on limited and noisy data. 

Returning to our example system, we can say that Eq.~(\ref{eqexjac}) captures the dynamics around all steady states in all models of the form of Eq.~\ref{eqGMex} by three intuitive parameters: the elasticity of gain and loss, and a turnover rate. 

A steady state is stable if all eigenvalues of the Jacobian have negative real parts. In our one-dimensional example the Jacobian is a $1\times 1$ matrix and thus has only one eigenvalue which is identical to the matrix element itself, 
\eq{
\lambda = \alpha (g_{\rm x} - l_{\rm x}).
}
Hence, a steady state under consideration is stable (i.e.~$\lambda<0$) if
\eq{
g_{\rm x}<l_{\rm x}.
}
Even in this very simple example the analysis reveals a concrete result. For all steady states in all systems where a single positive variable is subject to gains and losses (systems of the form of Eq.~(\ref{eqGMex})) the following is true:
\begin{itemize}
    \item Turnover rate does not directly impact stability. (It may however indirectly impact stability, for example if losses experience stronger nonlinearity under increased turnover.)
    \item A steady state is stable if the elasticity of loss is greater than the elasticity of gain in the steady state.
\end{itemize}
To readers with experience in modeling, these results will be hardly surprising. Consider however, that in this small example we have given the GM only very little structural information to work with. We show below that the same procedure can be applied to models of almost arbitrary complexity. If we provide more information, e.g.~by specifying complex food-web topologies or metabolic networks, GM can reveal deeper, more detailed insights.  

\section{Frequently asked questions}
Our introduction to GM continues in Sec.~\ref{secProcedure}, below. However, in our experience researchers frequently have specific questions after seeing the first introductory example. We therefore seize on this opportunity to answer the most common ones in this frequently asked questions section.\vspace{2mm} 

\noindent{\bf What if there are multiple steady states?\\}
The general model captures the stability of all of them, but because the normalization is done with respect to the steady state, different steady states will be described by different parameter values.\vspace{2mm}

\noindent{\bf It seems too easy. Does it actually work?\\}
Yes it does, the procedure, as it it has been spelled out here has been supported by rigorous mathematical proofs \citep{Kuehn2013AppMath}. Perhaps more importantly the papers cited in Sec.2 provide plenty of evidence that valuable information can be gained by GM.\vspace{2mm}

\noindent{\bf Are there some things you can't do with generalized models? Why doesn't everybody use them?\\}
There are a lot of things that cannot be done with GM, for example you can never compute where a steady state actually is. Also, you cannot simulate a GM, but you can explore it with other analysis methods that are safer, more efficient and often more powerful than simulation. GMs are not meant to replace conventional modeling, instead they are an additional tool by which some information can be gained very cleanly and efficiently. In practice they are used to explore model structures and to narrow down on parameter regions that are feasible, plausible and interesting before exploring in more detail with conventional models. \vspace{2mm}

\noindent{\bf Is this limited to analysis of steady states?\\}
In principle, no, in practice mostly yes. Christian K\"uhn has developed a method by which GMs can be used to explore the stability of other attractors \cite{Kuehn2013DCDS}. While mathematically sound, this extension requires to parameterize the shape of the attractor, which increases the size of the parameter space. Moreover instead of the comparatively simple stability analysis we have to analyze the model using Floquet theory. In practice an easier alternative is often to change the way the system is modelled. For example instead of modeling a differential equation system that has a limit cycle we can directly model the Poincar'e map, in which the cycle appears as fixed point and can hence be explored by local analysis. \vspace{2mm}

\noindent{\bf Is this only useful for stability analysis?\\}
Kind of, but not entirely. Stability analysis and its downstream products (Bifurcations, Robustness, Identification of sensitive or influential variables) are currently the best tools in the GM toolbox. But some other analysis can be done as well (see Sec.~\ref{secAnalysis}). \vspace{2mm}

\noindent{\bf Can I have processes that become negative?\\}
No, if your processes become negative, it breaks the normalization. In practice, there is an easy fix to this: Instead of having a process that can run in both directions define two processes that run antagonistically. For example, in a chemical reaction we would treat forward and reverse reactions as two different processes, which may be differently regulated. When modeling ecological dispersal between habitat patches, we model emigration as a separate process from immigration. In most cases this leads to better and more interpretable models. \vspace{2mm}

\noindent{\bf How about variables or processes becoming zero?\\}In general this is not a big problem. Suppose you have an ecological model in which some species can go extinct. If you analyze the full GM of the system then the results will apply to steady states where all species coexist. Also, we can verify that the steady states under consideration is a state where all species are present as this information is reflected in the generalized parameters. If we are particularly interested in the case where one of the species goes extinct then we can make another model where that species is absent. We might also be interested in the transition where the extinction occurs, and in general the GM can find it. Consider that the model in which the species is present remains valid as we approach the point of extinction. Validity in this limit is sufficient to detect the transition in which the extinction occurs. \vspace{2mm}  

\noindent{\bf What about conversation laws? Other peculiarities of my system?\\}
Conservation laws present us with similar issues as the stationarity conditions discussed above. Such additional constraints provide an opportunity to narrow down the parameters space. However, we must ensure that we seize this opportunity to avoid parameterizing systems that violate the conservation law and hence cannot exist in the real world. In the context of GMs conservation laws and also some other knowledge that we might have about the real system can be taken into account in the form of algebraic equation. This is described in the next section.  


\section{General modeling procedure} \label{secProcedure}
We now turn to the procedure by which complex GMs are formulated. For this purpose we follow the example of a simple predator-prey system from \cite{Yeakel2011TE}, but discuss it in greater detail.

\subsection{Identification of state variables}
The first step in this process is to identify the state variables that we want to describe. Deciding what state variable should be included is often easy, but can become complicated in models including human behavior. 

In GM introducing additional state variables often pays off in terms of interpretability and does not incur a high cost in terms of tractability. Hence, it is generally advisable to include a candidate variable rather than leaving it out. This will typically lead to large but sparse Jacobians that are often preferable to small dense ones.        

\subsection{Identification of processes}
Once our state variables are in place we have to identify the processes that change them. Each state variable needs at least one gain and one loss process, but there can be multiple of these processes. For example a simple predator-prey model could look like this 
\eqa{
\dot{X}&=&S(X)-F(X,Y)-L(X)\\
\dot{Y}&=&G(X,Y)-M(Y)
}
where $S$ describes the reproduction of the prey $X$, $F$ is the loss of prey due to predation, $L(X)$ is the loss of prey due to other causes than predation, $G$ is the gain by predation of predator $Y$, and $M$ is the predator's mortality. 

At this point we could ask why it is necessary to include for example the $L$ term at all. After all, since $S$ and $F$ are general functions we could easily merge $L$ into $F$ summarizing all the losses, or we could even merge $L$ into $S$ forming an effective growth term (as for example in logistic growth). However, in GM we extract insights mainly from the structure of the model. The more detailed we can specify the structure the more insights we gain. In the example merging $S$ and $L$ into one term would be a bad idea, because the exponent parameter of the merged term is far less interpretable than the two individual exponent parameters of $S$ and $L$. 

Similarly merging $F$ and $L$ would be a bad idea. As this is a predator-prey model we probably want to discuss predation losses separately from other losses. Splitting the processes preserves this ability. Moreover, it allows us further elaboration of the relationship between $F$ and $G$, shown below.

\subsection{Normalization}
We now define normalized variables and processes following the same procedure as in the introductory example, i.e.~for every variable $X$ we define 
\eq{
x=\frac{X}{X^*},
}
where $X^*$ is an unspecified stationary state, and for every process $P(X)$ we define
\eq{
p(x)=\frac{P(xX^*)}{P^*},
}
where $P^*=P(X^*)$. Processes of multiple variables can be dealt with analogously. For our example system this leads to the normalized differential equations 
\eqa{
\dot{x}&=&\frac{S^*}{X^*}s(x)-\frac{F^*}{X^*}f(x,y)-\frac{L^*}{X^*}l(x)\\
\dot{y}&=&\frac{G^*}{Y^*}g(x,y)-\frac{M^*}{Y^*}m(y)
}
Our next goal is to capture the prefactors in meaningful scale parameters while taking the stationarity condition of the steady state into account. Although there is some freedom in the way we want to specify our scale parameters, it is often a good idea to use one parameter per variable to denote the total turnover and then define additional parameters as needed to specify how much the individual gains and losses contribute to the turnover. 

To introduce parameters in an organized fashion we proceed as follows: We start by considering our normalized differential equations in the steady state, where the time derivative vanishes and all process rates are 1,
\eqa{
0&=&\frac{S^*}{X^*}-\frac{F^*}{X^*}-\frac{L^*}{X^*}\\
0&=&\frac{G^*}{Y^*}-\frac{M^*}{Y^*}
}
These equations state that for each of the variables the sum of the loss terms is identical to the sum of the gain terms,
\eqa{
\frac{S^*}{X^*}&=&\frac{F^*}{X^*}+\frac{L^*}{X^*}\\
\frac{G^*}{Y^*}&=&\frac{M^*}{Y^*}
}
hence we define 
\eqa{
\alpha_{\rm x}&=&\frac{S^*}{X^*}=\frac{F^*}{X^*}+\frac{L^*}{X^*} \\\label{eqalphay}
\alpha_{\rm y}&=&\frac{G^*}{Y^*}=\frac{M^*}{Y^*}.
}
Where $\alpha_{\rm x}$ and $\alpha_{\rm y}$ are now our turnover parameters for the two species. 
Using these parameters, we can now write our differential equations as 
\eqa{
\dot{x}&=&\alpha_{\rm x}(s(x)-\frac{1}{\alpha_{\rm x}}\frac{F^*}{X^*}f(x,y)-\frac{1}{\alpha_{\rm x}}\frac{L^*}{X^*}l(x))\\
\dot{y}&=&\alpha_{\rm y}(g(x,y)-m(y)).
}
We are almost done here, but we still need to take care of the prefactors in front of the $f$ and $l$ terms. Note that by pulling the turnover rates out of these factors we created some interesting expressions. Let's use $\rho$ to denote the factor in front of the $f$-term. We can write   
\eq{
\rho = \frac{1}{\alpha_{\rm x}}\frac{F^*}{X^*} = \frac{F^*}{F^*+L^*}, 
}
which shows that $\rho$ is the proportion of the prey's loss in the steady state that is due to predation. Likewise, we can define
\eq{
\bar{\rho} 
= \frac{1}{\alpha_{\rm x}}\frac{L^*}{X^*} = \frac{L^*}{F^*+L^*},
}
which is the proportion of the loss in the steady state that occurs due to other sources of mortality. Such scale parameters that describe the branching or merging of flows within the system are called \emph{branching parameters}. When we introduce such parameters, we have to keep in mind that they are not independent as the branching parameters for the gains or losses of a particular variable always add up to one. In our example we can quickly verify that $\rho$ and $\bar{\rho}$ must obey
\eq{
\rho+\bar{\rho}=\frac{F^*}{F^*+L^*}+\frac{L^*}{F^*+L^*} = 1.
}
Keeping this constraint in mind we can now write our model as 
\eqa{
\dot{x}&=&\alpha_{\rm x}(s(x)-\rho f(x,y)-\bar{\rho}l(x))\\
\dot{y}&=&\alpha_{\rm y}(g(x,y)-m(y))\\
\bar{\rho}&=&1-\rho
}
In general, it will not be necessary to go through the normalization in such detail as the outcomes always follow the same pattern. We can thus quickly see that for example the equation 
\eq{
\dot{Z}=A(X)+B(Y)+C(Z)-Q(Z)-R(Z,Y)-S(Z,Y,X)
}
normalizes to 
\eqa{
\dot{z}&=&\alpha(\beta_{\rm a}a(x)+\beta_{\rm b}b(y)+\beta_{\rm c}c(z)-\sigma_{\rm q}q(z)-\sigma_{\rm r}r(z,y)-\sigma_{\rm s}s(z,y,x)) \\ 
1 &=& \beta_{\rm a} + \beta_{\rm b} + \beta_{\rm c} \\
1 &=& \sigma_{\rm q} + \sigma_{\rm r} + \sigma_{\rm s}.
}

\subsection{Timescale normalization and Jacobian}
Before we calculate the Jacobian, we can always remove one of our turnover parameters by rescaling time. For example, if we measure in terms of multiples of the turnover time of the prey $1/\alpha_{x}$ both equations are rescaled by this factor. As a result, we can write the system as 
\eqa{
\dot{x}&=&s(x)-\rho f(x,y)-\bar{\rho}l(x)\\
\dot{y}&=&\alpha(g(x,y)-m(y))\\
\bar{\rho}&=&1-\rho
}
where $\alpha=\alpha_{\rm y}/\alpha_{\rm x}$ is the parameter that tells us the relative rate of predator turnover to prey turnover. If the currency of our model is abundance this factor is the prey life expectancy divided by the predator life expectancy. If the currency of the model is biomass, the turnover ratio is the ratio of metabolic rates which is typically governed by allometric scaling laws \citep{yeakel2018dynamics}.  
 
To construct the Jacobian matrix of our predator-prey example we compute the derivatives 
\eqa{
\left.\frac{\partial \dot{x}}{\partial x}\right|_1 &=& s_{\rm x}-\rho f_{\rm x} - (1-\rho) l_{\rm x} \\
\left.\frac{\partial \dot{x}}{\partial y}\right|_1 &=& -\rho f_{\rm y} \\
\left.\frac{\partial \dot{y}}{\partial x}\right|_1 &=& \alpha g_{\rm x} \\
\left.\frac{\partial \dot{y}}{\partial y}\right|_1 &=& \alpha (g_{\rm y}-m_{\rm y}) \\
}
where we defined the exponent parameters $s_{\rm x}, f_{\rm x}, f_{\rm y}, l_{\rm x}, g_{\rm x}, g_{\rm y}, m_{\rm y}$ as needed and substituted $\bar{\rho}=1-\rho$. We can now write the Jacobian as 
\eq{
{\bf J} = \avecc{ s_{\rm x}-\rho f_{\rm x} - (1-\rho) l_{\rm x} & -\rho f_{\rm y} \\ \alpha g_{\rm x} & \alpha (g_{\rm y}-m_{\rm y}) }.
}
Compared to Jacobians that we typically find in conventional models this is a relatively simple and neat matrix, nevertheless it captures many insights into the structure of the system. Note in particular that we were not forced to make assumptions on aspects of the system that we are typically uncertain about, such as the exact form of predator-prey kinetics. By contrast many structural features that we can be certain about are represented. For example, net growth is the differences between gains and losses, independent processes add up, prey reproduction is assumed to be independent of the predator, etc. 

\subsection{Additional constraints and auxiliary variables}
In our example system the gain of the predator is still disconnected from the loss of the prey. Surely the functions $F$ and $G$ in our original model are not independent. But they are also not necessarily identical. As a basic approach we could assume that the predator gain is a function of the prey loss, e.g.~
\eq{
G(X,Y)=H(F(X,Y))
}
At this point one may wonder if it is useful at all to write a relationship where an unspecified function $G$ depends on an unspecified function $F$ in an unspecified way $H$. In GM the answer is generally yes, because all of these functions correspond to well defined elements of our mental model: predation loss $F$, conversion efficiency $H$, predation gain $G$.
By representing these elements separately, we make them tangible in the equations, or, in other words, the equations become a better representation of what we have in mind when we consider the system. 

Imposing such additional constraints on a GM, typically results in additional constraints on the scale and or exponent parameters. In this example we can quickly verify that there is no impact on the scale parameters: In the steady state our new condition just reads $G^*=H^*$ which does not impose any condition on existing scale parameters that would constrain their values. 

To find the implications of the condition for the exponent parameters we normalize the condition using the same procedure that we applied to the differential equations. In this case, we start by defining 
\eq{
h(f)=\frac{H(F^*f,Y^*y)}{H^*}.
}
Then we can start from the normalization of $G$ and write 
\eq{
g(x,y) = \frac{G(X^*x,Y^*y)}{G^*} = \frac{H(F^*f,Y^*y)}{H^*} = h(f(x,y)). 
}
We can then compute the exponent parameters 
\eqa{
g_{\rm x} &=& \left.\frac{\partial}{\partial x} h(f(x,y))\right|_1 = h_{\rm f}f_{\rm x} \\ 
g_{\rm y} &=& \left.\frac{\partial}{\partial y} h(f(x,y))\right|_1 = h_{\rm f}f_{\rm y} 
}
These equations now fix two of our exponent parameters, while a new parameter $h_{\rm f}$ appears that captures the elasticity of predator gain with respect to prey loss. If we are willing to assume a constant conversion efficiency as most models do $H$ is a linear function and hence $h_{\rm f}=1$. 

The biomass conversion example presented here is still a very basic case. In other papers the same approach has been used to build significantly more complex relationships into GM. For example \cite{Gross2006PRE} show such auxiliary constraints can be used to build realistic prey-switching behavior into GMs. 

For an intermediate illustration, let us think a bit deeper about biomass conversion. Ecological intuition suggests that conversion efficiency should depend on the per-capita consumption of prey by predators. Building this ecological insight into the GM gives us more structure to work with and hence offers potentially deeper insights. So, let's consider an alternative version of $H$: 
\eq{
G(X,Y)=H(F,C)
}
where $C$ is the per-capita consumption 
\eq{
C(F,Y)=\frac{F(X,Y)}{Y}.
}
Even this form of the constraint does not constrain our scale parameters further. To find the constraints on the exponent parameters we define normalized forms of the auxiliary variables
\eq{
h(f,c)=\frac{H(F^*f,C^*c)}{H^*}\quad\quad\quad c(f,y)=\frac{C(F^*f,Y^*y)}{C^*}
}
and then verify
\eq{
g(x,y)=\frac{G(xX^*,yY^*)}{G^*}=\frac{H(fF^*,cC^*)}{H^*} = h(f,c)
}
We can now compute the derivatives, keeping in mind that the auxiliary variables ($h$, $c$) are just short hand notations that need to be differentiated using the chain rule
\eqa{
g_{\rm x} &=& h_{\rm f}f_{\rm x} + h_{\rm c}c_{\rm f}f_{\rm x} \\
g_{\rm y} &=& h_{\rm f}f_{\rm y} + h_{\rm c}(c_{\rm f}f_{\rm y}+c_{\rm y}) 
}
In the solution we interpret the exponent parameters as partial derivatives such that for example $c_y$ denotes only the derivative of $c$ with respect to the second argument, the indirect impact of $y$ on $c$ via $f$ is accounted for in the independent term $c_{\rm f}f_{\rm y}$. 

Let's try to interpret the parameters that appear here, $h_{\rm f}$ is the partial derivative of $h$ with respect to $x$, and since $c$ now appears as an explicit argument of $h$ it means that this is a derivative at constant $c$. So ecologically speaking this parameter is asking how does the growth of the predator population change if more predators are feeding but the per capita amount stays constant. This is almost certainly a linear relationship so we can assume $h_{\rm f}=1$ with much better confidence than before. The parameter $h_{\rm c}$ corresponds to the question what happens if every predator consumes a greater amount per capita. In this case the efficiency of conversion can go up or down in complex nonlinear ways depending on the ecological situation, so we keep this as a tunable parameter, whose effect can be explored with the GM. Finally, $c_{\rm f}$ and $c_{\rm y}$ describe the elasticities of the per-capita consumption. Since we specified $C$ explicitly we can compute 
\eq{
c_{\rm f} = \left. \frac{\partial}{\partial f} c(f,y) \right|_1  
         = \left. \frac{\partial}{\partial f} \frac{C(F^*f,Y^*y)}{C(F^*,Y^*)} \right|_1
         = \left. \frac{\partial}{\partial f} \frac{F^*f}{Y^*y}\frac{Y^*}{F^*} \right|_1
         =  \left. \frac{\partial}{\partial f} \frac{f}{y} \right|_1
         =  1,  
} 
which we could have guessed straight away because we defined $C$ to be linear in $F$. Similarly,
\eq{
c_{\rm y}=  \left. \frac{\partial}{\partial y} \frac{f}{y} \right|_1
         =  -1,  
}
which is consistent with the inverse relationship we assumed. Summarizing these calculations we can write 
\eqa{
g_{\rm x} &=& (1 + h_{\rm c})f_{\rm x} \\
g_{\rm y} &=& f_{\rm y} + h_{\rm c}(f_{\rm y}-1) 
}
We have included this slightly more difficult example in this review because we feel that it illustrates very well that GMs have the ability to incorporate information that we are confident about (e.g.~$h_{\rm f}=1$) while not forcing us to constrain our options where we do not have such information (e.g.~$h_{\rm c}$). 

We can now substitute the results into the Jacobian matrix, which would remove the now redundant parameters $g_{\rm x}$ and $g_{\rm y}$ but insert the elasticity of conversion efficiency with respect to per capita consumption $H_{\rm c}$. We could then for example explore how this parameter impacts the stability of the system, or changing it affects the predator-prey ratio (see Sec.~\ref{secImpact})
 
\subsection{Conservation laws}
Conservation laws can be imposed on a GM in a similar way as the conditions, discussed above. However, there are two different ways in which a conservation law can be used, leading to slightly different notions of stability. 

For the purpose of illustration consider that we have a two-dimensional system subject to one conservation law. We could then use the conservation law to reduce the number of variables to one even before normalization. Subsequently the normalization can be carried out normally. Alternatively, we can normalize the system and the conservation law and then interpret the conservation law as a constraint on the generalized parameters. 

Both of these approaches are valid, but in the first case we arrive at a Jacobian of size $1\times 1$, whereas in the latter case we arrive at a Jacobian of size $2\times 2$. Both of these Jacobians describe the stability of steady states in the system, but in the former case only perturbations that respect the conservation law are allowed (hence the one-dimensional eigenspace), whereas in the latter also those perturbations are considered that violate the conservation law, leading to a slightly stricter notion of stability.  

For most systems the second alternative provides the better notion of stability, unless the system is fundamentally closed and no perturbation from the outside is imaginable. The second alternative is also often simpler to implement as the simpler internal structure more than compensates for the slightly larger size of the Jacobian. 

Even if we take the second route, conservation laws will impose some constraints on scale parameters, which is normally harmless, but can become complicated if we have to deal with many such constraints.
This happens for example in metabolic models where the number of atoms of different elements are conserved. In such a case some additional machinery is needed to help us manage our scale parameters. A convenient solution is to represent the scale parameters as a linear combination of a set of fundamental flux modes that obey all constraints. This is explained in detail in \cite{Steuer2007BI}.   

\subsection{Derivative conditions and optimality}
As a final remark in this section let us briefly mention that we can also impose additional constraints in the form of derivatives. This makes no sense in the context of our ecological example, but, abstractly speaking, we could demand  
\eq{
    \left.\frac{\partial P(X)}{\partial X} \right|_* = 0, 
}
where $P$ is some process and $X$ is some state variable. Again, we can normalize
\eq{
  0 =  \left. \frac{\partial P(X)}{\partial X} \right|_* =  \left. \frac{\partial P^*p(x)}{\partial x}\frac{\partial x}{\partial X} \right|_* = \frac{P^*}{X^*} p_{\rm x}  
}
This illustrates that sometimes scale parameters such as $P^*/X^*$ can appear in the normalization of additional constraints. They are often easily dealt with as they can typically be replaced by already defined symbols (e.g.~$\alpha_{\rm x}$). In this example the appearance of the scale parameter is of no consequence as the outcome stipulates $p_{\rm x}=0$, fixing one of the exponent parameter. 

The ability to specify conditions on the derivatives is interesting because it provides us with a way to demand that a variable must be in a local maximum or minimum of some function. This is useful for example if we study biological evolution in an adaptive dynamics model \citep{allen2013adaptive} and want to force the species to remain in locally evolutionary stable states. The same approach can also be used in governance or cooperation models to demand that agents allocate their resources optimally.   


\section{Analyzing Generalized Models}\label{secAnalysis}
The analysis of GMs is fundamentally more constrained than the analysis of conventional models as we lack the ability to simulate the model or compute the steady states. Nevertheless, GMs can be analyzed in a variety of ways. 

\begin{figure}
    \centering
    \includegraphics[width=\textwidth]{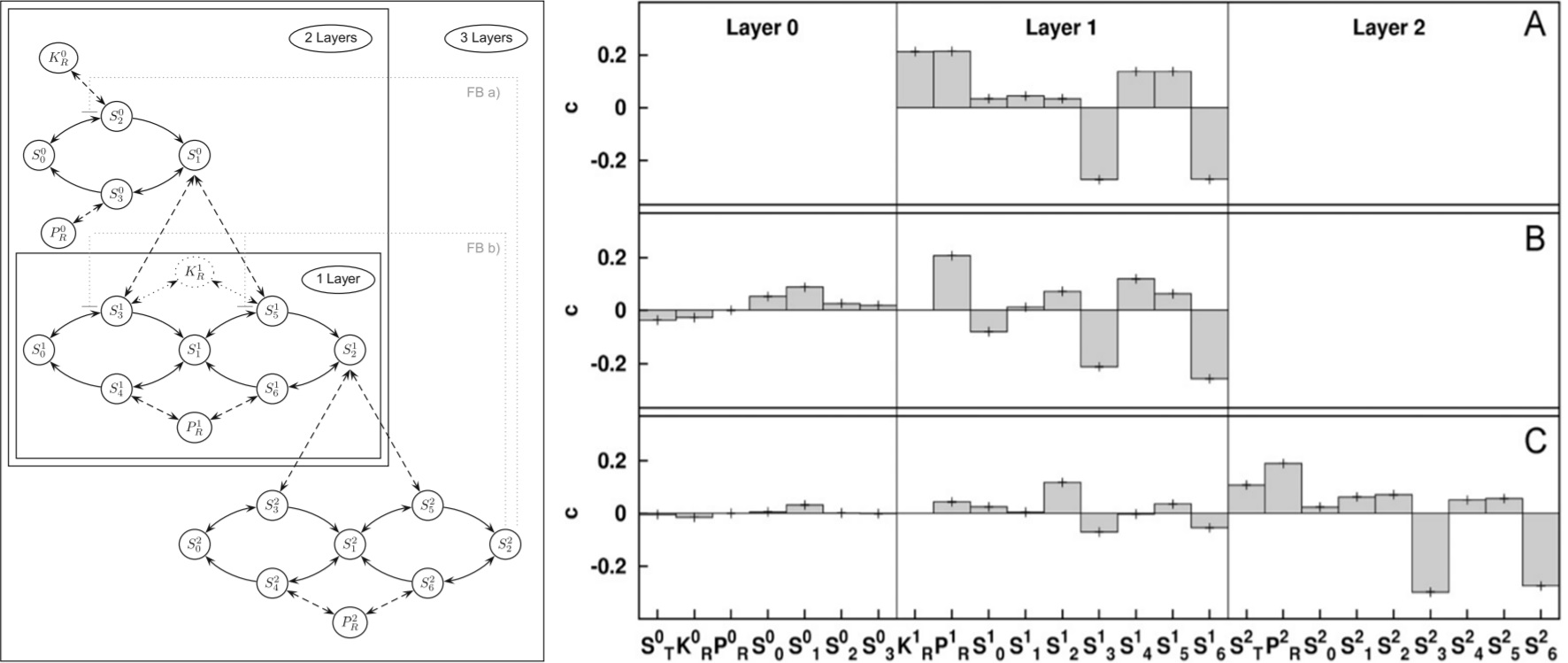}
    \caption{Illustration of stability sampling. The map kinase cascade (left) is a motif that appears in gene regulation. Numerical stability analysis of a GM reveals the impact different generalized parameters have on stability of different versions of the cascade that occur in biology: single layer (A), double layer (B), and triple layer (C). The results reveal stabilizing (positive) and destabilizing (negative) parameters. This numerical result is based on the analysis of 10 million steady states, which reduces the error bars to less than the line width of the plot. Using GM this number of states can be analyzed in seconds on modern computers. Figure adapted from \cite{Zumsande2010JTB}}.
    \label{figzumsande}
\end{figure}

\subsection{Numerical stability analysis} 
The main output of GM are Jacobian matrices. Hence the analysis of GMs is generally based on the analysis of Jacobians. The most direct application of these matrices is stability analysis. Given a specific set of generalized parameters, we can substitute the parameters into the Jacobian and then check the stability of the corresponding steady state by numerically computing the leading eigenvalue. 

Leading eigenvalues can be computed highly efficiently using iterative eigensolvers. We can take advantage of this efficiency to explore the parameter space spanned by the generalized parameters by random sampling. For this purpose, we constrain all parameters to plausible ranges, which is possible as the parameters are generally easy to interpret. For example we may decide to set the parameter $f_{\rm x}$ to cover the whole range from constant to quadratic response, i.e.~the same range of possibilities that would be covered by a Holling type-III functional response in conventional models \citep{holling1959some}. 

Once every parameter is thus constrained we can draw an ensemble of random data sets and evaluate their stability. We can then get a first impression of the behavior of a system by correlating the individual parameters with a stability. Suppose we have a system with $P$ generalized parameters and we draw $M$ sets of values for these parameters. We can then denote the $m$'th realization of parameter $i$ as $p_i^m$ where $i\in [1,P]$ and $m \in [1,M]$. Furthermore, we can denote the stability of the steady state described by the parameter set $m$ as $s_m$. 

It is tempting to define $s_m$ as $-{\rm Re}(\lambda_0)$, where $\lambda_0$ is the leading eigenvalue of the Jacobian. However, this can be misleading in large networks as instabilities can often arise on very different timescales, such that the eigenvalue that is the leading one in most of the parameter space is not the one that causes the instability once stability is lost. 

Instead we can define
\eq{
s_m = \left\{ \begin{array}{l l} 
  1 & \quad\quad\quad {\rm Re}(\lambda_0) < 0 \\
  0 & \quad\quad\quad \mbox{otherwise} 
  \end{array}\right.
}
so $s_m$ is one if the parameter set $m$ is stable and zero otherwise. Once we know the stability of all parameter sets we can estimate the impact of the individual parameters on stability as 
\eq{
c_i = {\rm Cov}_m(p_i^m,s_m)
}
If stability is completely explained by a particular parameter the result will be $c_i=1$ if the parameter is stabilizing or $c_i=-1$ if it is destabilizing. In general, many parameters will correlate with stability and typical values of important parameters in large networks are around 0.1 (see Fig.~\ref{figzumsande}). 

Let us emphasize that these correlations are not independent of the sampling. Sampling generalized parameters uniformly often results in a sensible sampling of the parameter space. Nevertheless, sampling parameters from wider ranges will result in stronger correlations. Therefore, it is essential to choose the ranges such that they reflect reasonable assumptions about the plausible values of the parameters.

Once we have identified a set of parameters of particular interest we can explore the effect of these parameters on stability in more detail. A common procedure is to vary one parameter $x$ systematically, while all other parameters are randomized. We can then plot the proportion of randomly drawn parameter sets that are stable over the value $x$. For historical reasons this measure is commonly called the \emph{probability of stable webs}. The same procedure can also be used with two parameters to produce two-dimensional histograms (Fig.~\ref{figLarsplot}). 

\begin{figure}[ht]
    \centering
    \includegraphics[width=\textwidth]{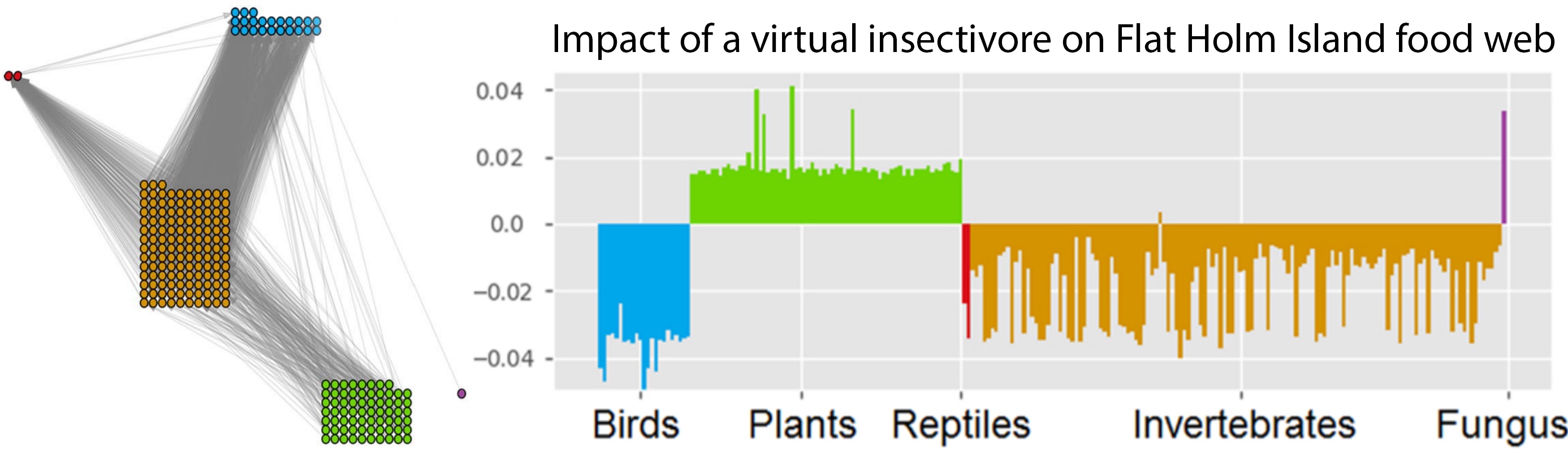}
    \caption{Impact of a (virtual) insectivore on the Flat Holm island food web. The impact is assessed with a GM of the empirically observed food web (left). This network contains 227 species (23 birds, 2 reptiles, 133 invertebrates, 68 plants and 1 fungus). GM impact analysis predicts which species would profit and which species would suffer from a typical insectivore invasion (right). This reveals a general pattern of responses but can also be used to identify indicator species for the detection of bioinvasions. Figure adapted from \cite{doizy2018impact}. 
    }
    \label{figcyberinvasive}
\end{figure}

\subsection{Response to parameter change \label{secImpact}}
In addition to stability against short-term (pulse) perturbations, we can also ask how a dynamical system responds to a permanent change of parameters, i.e.~a press perturbation. 

By the implicit function theorem one can show (see \citet{Aufderheide2013PNAS}) that a sufficiently small press perturbations induces a shift in the steady states described by   
\eq{
 \label{eqPerturb}
\vec{\delta} 
= -{\mat{J}^{-1}}\vec{p}.
}
where $\vec{delta}$ is the induced shift in the steady state,
${\bf J}^{-1}$ is the inverse of the Jacobian matrix, and $\vec{p}$
is a vector containing the direct impacts of the perturbations on the individual equations. 

If we apply this equation to a GM then the vector $\vec{\delta}$ will contain the impact on the steady state in the normalized variables, e.g.~an element of 0.05 would correspond to a 5\% increase. Similarly, $\vec{p}$ contains the direct impact on the differential equations in units of normalized turnover.   

For example let us again consider the predator-prey system from Sec.~6. We can now ask, how switching on an additional loss term for the prey, say, from harvesting, impacts the steady state. We assume that a small fraction $\epsilon$ of the total turnover of the prey is harvested, hence
\eq{
\vec{p} = \avec{\epsilon \\ 0}.
}
Substituting this relationship and the Jacobian into Eq.~(\ref{eqPerturb}) we arrive at 
\eq{
\vec{\delta}
= -\frac{1}{\rm{det} \ \mat{J}} 
\avecc{ \alpha (g_{\rm y}-m_{\rm y}) & \rho f_{\rm y} \\ -\alpha g_{\rm x} &  s_{\rm x}-\rho f_{\rm x} - (1-\rho) l_{\rm x}}
\avecc{-\epsilon\\ 0},
}
where
\eq{
\rm{det}\ \mat{J} = \alpha((s_{\rm x}-\rho f_{\rm x}-(1-\rho) l_{\rm x}) (g_{\rm y}-m_{\rm y}) + g_{\rm x} \rho f_{\rm y}),
}
We can now examine the two components of the resulting impact on the steady state
\eq{
\delta_{\rm prey} 
= \frac{ \alpha(g_{\rm y} - m_{\rm y})}{\rm{det}} \epsilon , \quad\quad\quad
\delta_{\rm pred} 
= -\frac{ \alpha g_{\rm x}} {\rm{det}\ \mat{J}} \epsilon .
}
If there is no strong interference or social interaction between predators we can assume that predation is linear in predator abundance, $g_{\rm y} = 1$. Moreover in the absence of diseases and overcrowding the mortality of the predator can be assumed to be linear $m_{\rm y} = 1$. We can now see that in this case 
\eq{
\delta_{\rm prey} = \frac{ \alpha(g_{\rm y} - m_{\rm y})}{\rm{det}} \epsilon = 0
}
So the prey is not impacted by a small amount of harvesting at all. This happens because the prey population is still controlled by the predator and any additional losses of the prey are in the long run compensated by reduced predation. Even though the prey abundance does not change, we can see from $\delta_{\rm pred}$ that the predator population is impacted by the harvesting of its prey, and at low harvesting rates responds with a proportional loss. An example for GM impact analysis in a larger network is shown in Fig.~\ref{figcyberinvasive}.

\begin{figure}
    \centering
    \includegraphics[width=0.6\textwidth]{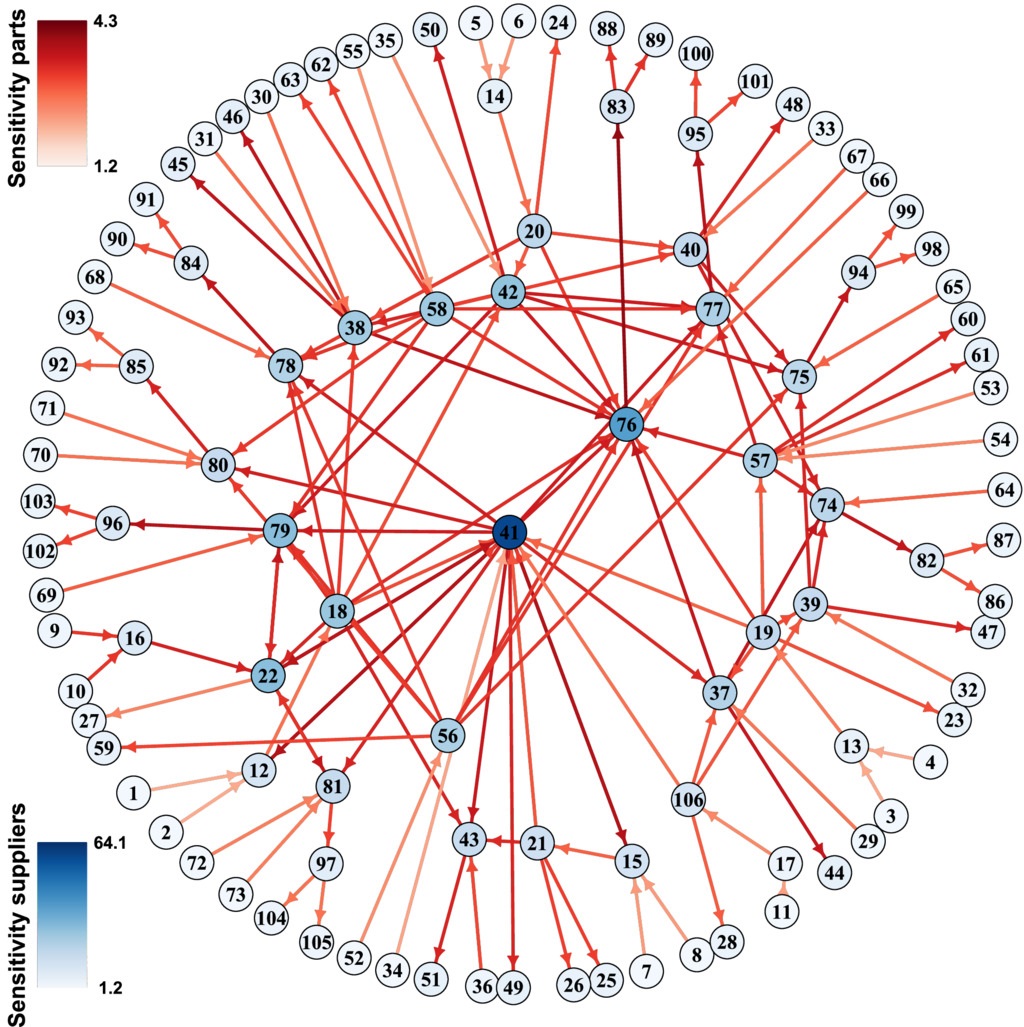}
    \caption{Sensitivity in a luxury goods supply network. GM was used to study the sensitivity (colors) of firms (nodes) and products (links) in a luxury goods supply network. Figure adapted from \cite{demirel2019identifying}.}
    \label{figsupply}
\end{figure}

In some applications we are not interested in the impact of a specific perturbation, but in the response to general perturbations. For this purpose \citet{Aufderheide2013PNAS} propose two measures
\eq{
{\rm{Se}_i} = {\rm {log}} \left(- \sum_{n} \frac{|v_{ i}|^{(n)}}{\lambda _{n}} \right), \quad\quad\quad
{\rm{In}_i} = {\rm {log}} \left(- \sum_{n} \frac{|w _{i}|^{(n)}}{\lambda _{n}}\right),
}
where $\lambda_n$ is the $n$th eigenvalue of $\bf J$, and $\vec{v_n}$ and $\vec{w_n}$ are the corresponding left and right eigenvectors. For example, $w_{1}|^{(2)}$ is the first element in the right eigenvector of $\bf J$ corresponding to the second eigenvalue.   

Among the two measures Se$_i$ indicates how sensitive variable $i$ will react to typical perturbations of the network, whereas Im$_i$ indicates how strongly a perturbation of node $i$ will propagate to typical nodes in the network. This former is illustrated in a manufacturing network in Fig.~\ref{figsupply}. If desired dynamical importance of a variable can be defined as the product of these measures.

\begin{figure}
    \centering
    \includegraphics[width=\textwidth]{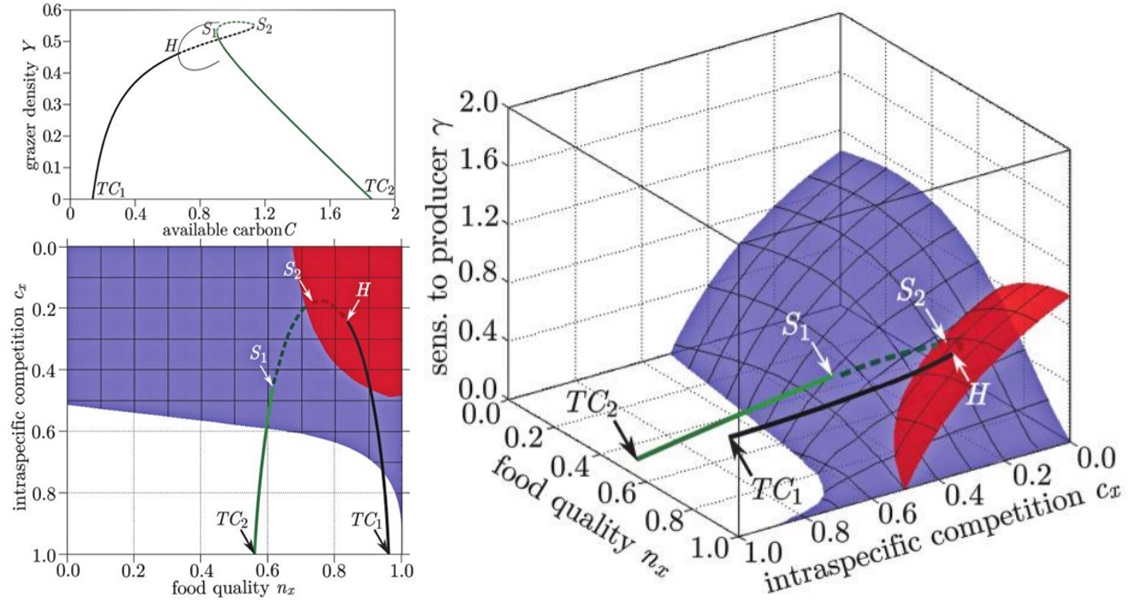}
    \caption{Comparison of bifurcation diagrams in conventional and generalized studies of food quality in a producer-grazer system. In the conventional model (top left) the steady state in which grazers survive emerges from a transcritical bifurcation (TC$_1$), as a conventional parameter is increased the steady state loses stability in a Hopf (H) bifurcation, where a limit cycle (thin line) is created. The steady state subsequently reacquires stability by going through two saddle-node bifurcations (S$_1$, S$_2$), before vanishing in another transcritical bifurcation (TC$_2$). The stability of all steady states is captured by the GM (shown in two projections: right, bottom left). In the GM the bifurcation points form surfaces (red: Hopf, blue saddle node) and the states visited by the conventional example model corresponds to a trajectory through the volume spanned by the GM. The surface on which the transcritical bifurcations occur is the identical to the front left face of space shown in the right diagram and is omitted to avoid occlusion. Figure adapted from \citet{Stiefs2010AMNAT}.}
    \label{figFood}
\end{figure}

\subsection{Mathematical stability analysis and bifurcation theory}
Because GM avoids the computation of steady states, the elements of the matrix are often simple expressions, which are convenient for pen and paper math. Hence GM is sometimes used to create Jacobians for methodological studies that need example Jacobians with certain properties. A recent example is \cite{barter2021closed} which proposed a method for inferring causality (in the form of the Jacobian matrix) from correlations and used GMs to create suitable examples. Other examples for this use of GMs include \cite{Hoefener2011EPL,Hoefener2012PTRS,Do2012NJP}.

The most common mathematical use of the Jacobian is bifurcation analysis, the study of the transitions between dynamical regimes in the system. In a system of differential equations local bifurcations of stationary states occur when a change in parameters causes at least one eigenvalue of the Jacobian to change sign \citep{Guckenheimer,Kuznetsov}. Crossing the threshold where such a bifurcation occurs typically leads to qualitative transition in the system.  

\begin{figure}
    \centering
    \includegraphics[width=\textwidth]{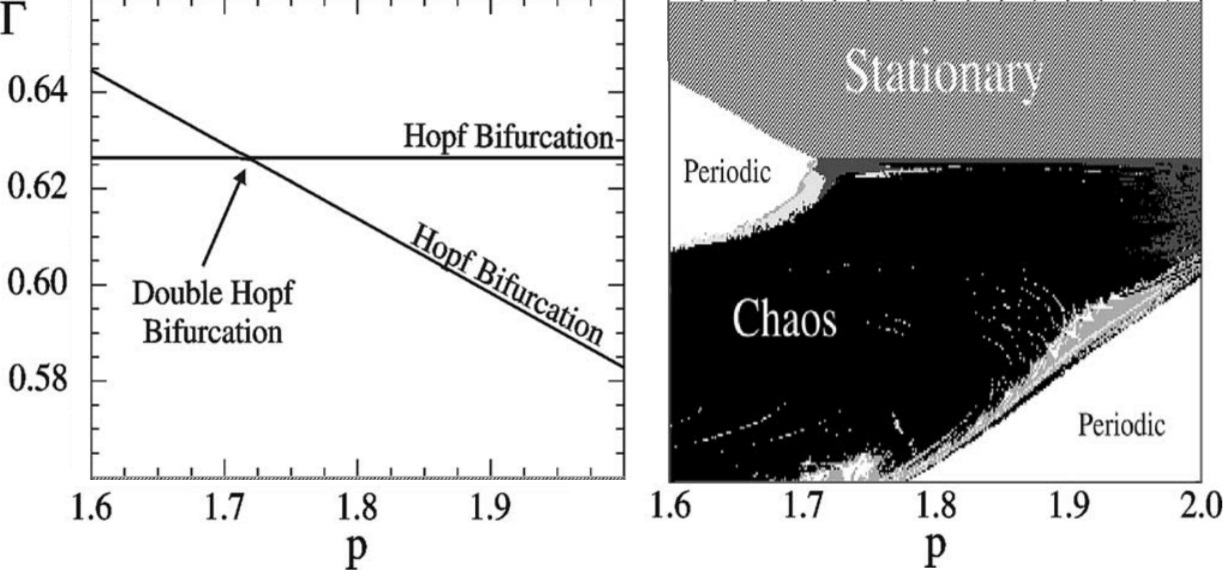}
    \caption{Chaos in a four-trophic food chain. A GM reveals that the steady states are stable in the topmost volume of the parameter space shown in the two-parameter bifurcation diagram (left). When parameters are changed, steady states can lose their stability by crossing either of two Hopf bifurcations (lines). At the intersection of these lines a codimension-2 double-Hopf bifurcation point is located. Observing this bifurcation in the GM points to a chaotic parameter region nearby. Indeed, this region can be discovered by numerical calculation of Lyapunov exponents in a conventional model (right). In the region where the GM predicts stability of the steady state the conventional model is stationary (dashed), in the unstable regions we observe oscillations (dark grey), quasi-periodicity (light and medium grey) and chaos (black). 
    Figure adapted from \cite{Gross2005OIKOS}.}
    \label{figchaos}
\end{figure}

There are only two generic ways in which bifurcations appear in GMs. Either a single real eigenvalue becomes zero or a complex conjugate eigenvalue pair crosses the imaginary axis (saddle-node bifurcation and its variants). The former scenario corresponds to bifurcations in which the number of steady states changes, whereas the latter typically marks the onset of oscillations (Hopf bifurcation) \citep{guckenheimer1997computing}.

The saddle-node-type bifurcations can be easily computed in GMs of any size. Here the task is to find the combinations of generalized parameters that lead to a zero eigenvalue of the Jacobian. Because the determinant of a matrix is the product of its eigenvalues \citep{Gantmacher} the Jacobian has a zero eigenvalue if and only if its determinant is zero. In contrast to the eigenvalues of a matrix which can only be computed for very small systems the determinant can be computed straight forwardly for systems of any size. 

For the Hopf bifurcation, we need to locate combinations of the generalized parameters that lead to a complex conjugate pair of purely imaginary eigenvalues. To find such pairs a determinant based method was proposed originally by \cite{guckenheimer1997computing} and rediscovered independently in \cite{Gross2004PD}. Using this method Hopf bifurcations can in principle be located in systems of any size, however the resulting equations become too complicated to be useful if the system has more than ca.~10 variables. 

Once we have located the bifurcations in a system, they can be visualized in bifurcation diagrams. In particular, several analysis of GMs have used three-parameter bifurcation diagrams, which can be created by the method described in \cite{Stiefs2008IJBC} (see Fig.~\ref{figFood}). Each point in the volume spanned by these diagrams corresponds to a particular steady state. All steady states located in the same volume share qualitatively similar local dynamical properties, whereas qualitative transitions take place at the surfaces, which mark bifurcation points. 

A de-facto convention in GM is to orient three-parameter diagrams such that the steady states that are stable are located in the top-most volume of parameter space, whereas the steady states in all other volumes are typically unstable. 

Three parameter bifurcation diagrams allow the researcher to quickly get an overview of the interaction of up to three relevant parameters. Moreover, they allow to quickly identify parameter regions where different bifurcations meet and intersect. This is particularly interesting because the intersections can reveal other dynamical features that are otherwise harder to discover. For example it is known that a region of chaotic dynamics must exist close to the intersection of two Hopf bifurcation surfaces \citep{Kuznetsov} (Fig.~\ref{figchaos}). This approach was used for example in \cite{zumsande2011general} to identify a region of chaotic dynamics in the MAP Kinase cascade, a common regulatory motif in cell biology.  

\subsection{Return to conventional models}
GMs are especially helpful in systems where large uncertainties exist. In these systems they can greatly speed up the initial exploration identifying interesting parameter regions and phenomena, based on limited information. However, as we progress beyond the initial exploration of the system we typically gain additional insights and/or data that we want to reflect in the model. Some of these additional insights can be used to restrict generalized parameter ranges. Others may lead to deeper understanding that results in changes to the model. 

One nice feature of GMs is that they can be iteratively expanded to reflect new insights into the system. 
For example in our initial exploration we may model a given process simply by a single unknown function $F(X,Y)$ and the resulting Jacobian will contain parameters such as $f_{\rm x}$ and $f_{\rm y}$. Once we understand this process in more detail we may want to replace this completely unconstrained function by a formulation that uses some newly gained insights. For example we may realize that $F$ is the product of some independent factors $F(X,Y)=A(X)Y$. We can now normalize this additional equation and use it to simply replace the old parameters in our Jacobian by new ones (in this example, $f_{\rm x}=a_{\rm x}$, $f_{\rm y}=1$).
In this way additional detail can be added to models iteratively without redoing the complete normalization procedure. This is discussed in more detail in \cite{Yeakel2011TE}.

As we understand our model better we may eventually want to restrict more and more functions to specific functional forms. This can lead to hybrid models where some equations are completely specified, whereas others still contain generalized terms. In this we need to solve manually for the steady states of fully specified equations, the resulting stationarity conditions then typically act as additional constraints on the generalized parameters. An example of such an hybrid model is the laser system discussed in \cite{Gross2006PRE}.  

Quite commonly, GM will identify parameter regions of particular interest. To explore the dynamics in these regions it is often desirable to run some numerical simulations. This means that we need a way to construct conventional models that are consistent with a given set of generalized parameters. In general, there will be many models that match the desired parameter set and many different ways to find them. However, the easiest and fastest procedure is to replace the general functions in the model by specific functions that obey the normalization condition $F(1)=\alpha$, where $\alpha$ is the desired turnover rate. This guarantees that the specific model that we are constructing still has a steady state at $X^*=1$ which saves us the work of computing the steady state. 

For illustration consider again our introductory example  
\eq{
\dot{X}=G(X)-L(X)
}
we know already that the Jacobian after timescale normalization is 
\eq{
{\bf J} = \alpha (g_{\rm x} - l_{\rm x})
}
Let's say we are interested in a steady state that is characterized by $\alpha=1$, $g_{\rm x}=1/2$ and $l_{\rm x}=2$. The challenge is now to find specific functions $G(X)$ and $L(X)$ such that there is a steady state that matches these parameter values. One class of functions that meet our condition $F(1)=1$ are the power laws $F(X)=X^p$. Computing the exponent parameter corresponding to such a power law, yields $p$. Thus, $G(X)=X^{1/2}$ and $L(X)=X^2$ meet the stationarity condition and match the desired parameters. Hence one possible example model is 
\eq{
\dot{X}=X^{1/2}-X^2
}
This is already a solution, but let's say the first term $X^{1/2}$ would be unrealistic in the context of our application. Instead we want a term of the form 
\eq{
G(X)= \frac{AX}{X+K}.
}
To make this work, we first enforce the normalization conditions $G(1)=1$, by setting $A=1+K$. Then we choose $K$ such that 
\eq{
p=\frac{1}{2} = \left. \frac{\partial}{\partial X}  \frac{(1+K)X}{K+X} \right|_1
= \frac{K}{(K+1)}  
}
which required $K=1$. Hence, also 
\eq{
\dot{X}= \frac{2X}{1+X} -X^2
}
is a specific model that is consistent with the desired parameter values. The same approach can be used to construct specific realizations of complex GMs. Therefore this method can serve as a constructive proof that each point in the generalized parameter space corresponds to a realizable steady state in a plausible conventional model \citep{Kuehn2013AppMath}. 

A small caveat regarding the procedure above is that the specific construction results in degeneracy in certain bifurcations. For simulation studies this is normally not a concern, but may cause peculiar results in bifurcation analyses.  

\section{Summary and Discussion}
In the present paper we summarized the state of the art in generalized modeling. In the past generalized modeling has been used in more than 50 publications in diverse areas of Science, Engineering, Mathematics and Medicine. As a result, the method has seen increasing adoption by labs around the world. 

Although generalized modeling is mathematically straight forward, its philosophy differs in important ways from conventional modeling. In our experience, these differences mean that young researchers with limited experience in dynamics find it easier to adopt generalized modeling than seasoned modelers, for whom it takes greater cognitive effort to switch to a different mental framework.   

Conventional modeling is very much grounded in the belief in an ultimate truth: At least within the model, the variables are governed by precise and exact rules and equations that given some initial conditions permit only one possible outcome. This is even true for stochastic models, which postulate micro-scale randomness, but work with precisely defined laws on the level of distributions. 

By contrast, generalized models, acknowledge that we have a limited view of reality and may hence be unable to perceive the exact laws that are at work in the system. Instead of postulating one definite reality, generalized models work with the whole infinite ensemble of possible realities that are consistent with the available structural knowledge. They explore dynamical implications within this ensemble, allowing the researcher to further narrow down the set of possible worlds. 

As we have shown, some questions can be answered very efficiently by analyzing the whole ensemble of possible worlds captured by the generalized model. These questions include, the analysis of dynamical stability and bifurcations of steady states, prediction of responses to different types of perturbations, and identification of important parameters and parameter regions. 

In summary generalized modeling offers a highly efficient approach to extract types of insights from limited information. This efficiency of generalized modeling is not limited to numerical efficiency, but also allows mathematical solutions in systems of intermediate complexity, and perhaps most-importantly saves the researcher time. Formulating a generalized model involves considerably less work than a comparable conventional model. It avoids extensive research and considerations which may be necessary in a conventional model to fix rate constants and parameterize kinetic laws.

Once a researcher is familiar with the general procedure and, more importantly, has adapted to its philosophy, generalized models can typically be formulated, analyzed and adapted within just a few hours. We hope that this review will help many new researchers to discover this exciting and entertaining modeling approach.  

\section*{Author Contributions}
The authors worked together on all aspects of this paper. 

\section*{Funding}
This work was supported by the Ministry of Science and Culture of Lower Saxony (HIFMB Project) and the Volkswagen Foundation (ZN3285).

\section*{Data Availability Statement}
This paper does not use any research data.

\end{document}